\documentclass[MSNbibl,number,dvips]{arxstspdf}
\usepackage{flushend}
\usepackage{stfloats}
\usepackage{graphicx}

\volume{29}
\issue{2}
\pubyear{2014}
\firstpage{214}
\lastpage{226}
\doi{10.1214/14-STS476} 


\begin{document}
\begin{frontmatter}

\title{Scalable Genomics with R and~Bioconductor}
\runtitle{Scalable genomics}

\begin{aug}
\author[a]{\fnms{Michael}~\snm{Lawrence}\corref{}\ead[label=e1]{michafla@gene.com}}
\and
\author[b]{\fnms{Martin}~\snm{Morgan}\ead[label=e2]{mtmorgan@fhcrc.org}}
\runauthor{M. Lawrence and M. Morgan}

\affiliation{Genentech and Fred Hutchinson Cancer Research Center}

\address[a]{Michael Lawrence is Computational Biologist, Genentech, 1 DNA Way, South San
Francisco, California 94080,
USA \printead{e1}.}
\address[b]{Martin Morgan is Principal Staff Scientist, Fred Hutchinson Cancer Research
Center, 1100 Fairview \mbox{Ave.~N.}, P.O. Box 19024, Seattle, Washington 98109,
USA \printead{e2}.}
\end{aug}

%
\begin{abstract}
This paper reviews strategies for solving problems encountered when
analyzing large genomic data sets and describes the implementation of
those strategies in R by packages from the Bioconductor project. We
treat the scalable processing, summarization and visualization of
big genomic data. The general ideas are well established and include
restrictive queries, compression, iteration and parallel
computing. We demonstrate the strategies by applying Bioconductor
packages to the detection and analysis of genetic variants from a
whole genome sequencing experiment.
\end{abstract}

%
\end{frontmatter}

\section{Introduction}
\label{sec-1}

Big data is encountered in genomics for two
reasons: the size of the genome and the heterogeneity of populations.
Complex organisms, such as plants and animals, have genomes on the
order of billions of base pairs (the human genome consists of over
three billion base pairs). The diversity of populations, whether of
organisms, tissues or cells, means we need to sample deeply to
detect low frequency events. To interrogate long and/or numerous
genomic sequences, many measurements are necessary. For example, a
typical whole genome sequencing experiment will consist of over one
billion reads of 75--100~bp each. The reads are aligned across
billions of positions, most of which have been annotated
in some way. This experiment may be repeated for thousands of
samples. Such a data set does not fit within the memory of a current
commodity computer, and is not processed in a timely and interactive
manner. To successfully wrangle a large data set, we need to
intimately understand its structure and carefully consider the
questions posed of it.

\begin{figure*}

\includegraphics{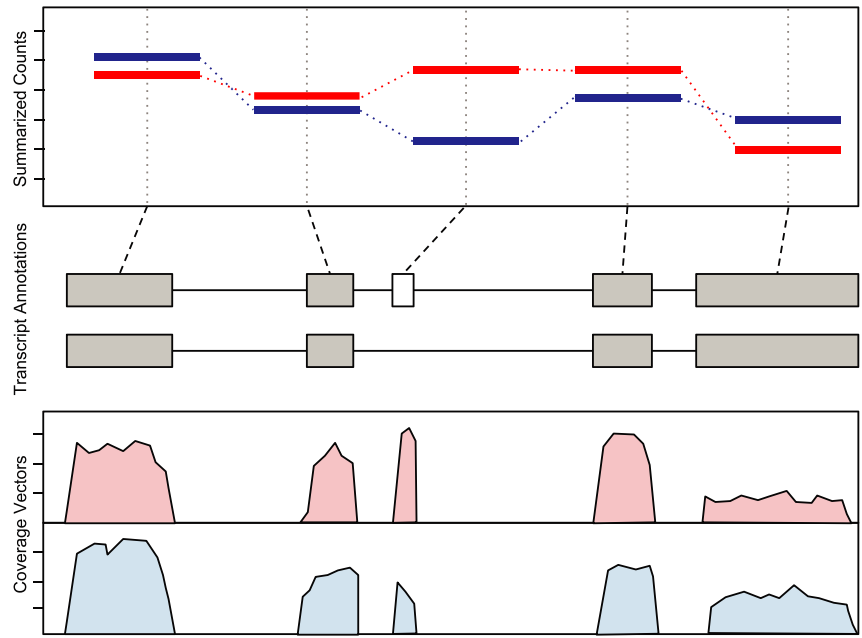}

\caption{Cartoon visualization of all three types of genomic data:
genome-length vectors, ranged, features and summaries (from bottom to
top). Data from two samples are compared (red vs. blue). The bottom
plot displays the coverage, a genome-length vector. The exon ranges are
shown in the middle. The top plot presents summaries, the per-exon read
counts.}\label{fig:three-data-types}
\end{figure*}

There are three primary types of data in genomics: sequence vectors,
annotated ranges and feature-by-sample matrices of summary
statistics (Figure~\ref{fig:three-data-types}).
Biological sequences are represented as strings of characters from a
restricted alphabet. For example, a DNA sequence consists of the
letters A, C, G and T, each referring to a particular type of
nucleotide. A genome consists of a set of DNA sequences, one for
each chromosome. We generalize the concept of sequence and define
the term \emph{sequence vector} to mean either a sequence or a vector
that runs parallel to a sequence. The latter may be some curated or
computed value, such as the cross-species conservation or coverage
from a sequencing experiment. The coverage is a common summary that
represents the number of features overlapping each position in the
reference sequence.

As we learn more about a genome, we annotate genomic ranges with
information like gene structures and regulatory binding sites. The
alignment of sequences to a reference genome is another type of
range-based annotation.

To compare data across samples, we often summarize experimental
annotations over a set of reference features to yield a
feature-by-sample matrix. For example, we might count read alignments
overlapping a common set of genes across a number of samples. Larger
matrices often arise in genetics, where thousands of samples are
compared over millions of SNPs, positions that are known to vary
within a population. In every case, the summaries are tied to a
genomic range.

To analyze the results of an experiment, we need to integrate data of
different types. For example, we might have alignments for a
ChIP-seq experiment, where the sequences have been enriched for
binding to a particular transcription factor. A typical analysis
involves checking coincidence with annotated binding sites for that
transcription factor, as well as looking for correlation between
gene expression and binding signal upstream of the gene. The gene
expression values might be drawn from a gene by a sample matrix
summarized from an RNA-seq experiment. The genomic range is the
common thread that integrates all three types of data. The sequence,
that is, the genome, acts as the scaffold, and ranges coordinate the
annotations and summarized features by locating them on the same
sequence.

The R language \cite{rman} is widely applied to problems in
statistics and data analysis, including the analysis of genomic
data \cite{pmid23950696}, as evidenced by the large number of available software
packages providing features ranging from data manipulation to
machine learning. R provides high-level
programming abstractions that make it accessible to statisticians
and bioinformatics professionals who are not software engineers per
se. One aspect of R that is particularly useful is its ``copy on
write'' memory semantics, which insulates the user from the
details of reference-based memory management.
The fundamental R data structure is the \emph{atomic vector}, which is
both convenient and efficient for moderately sized data. An atomic
vector is homogeneous in data type and so easily stored in one
contiguous block of memory. Many vector operations are implemented in native
(C) code, which avoids invoking the R interpreter as it iterates
over vector elements. In a typical multivariate data set, there
is heterogeneity in data type across the columns and homogeneity
along a column, so vectors are naturally suited for column-oriented
data storage, as in the basic \emph{data.frame}.
Vectorized computations can usually be expressed with simpler and more
concise code compared
to explicit iteration.
The strengths of R are also its weaknesses: the R API encourages
users to store entire data sets in memory as vectors. These vectors are
implicitly and silently copied to achieve copy-on-write semantics,
contributing to high memory usage and poor performance.

There are general strategies for handling large genomic data that
are well suited to R programs. Sometimes the analyst is only
interested in one aspect of the data, such as that overlapping a
single gene. In such cases, restricting the data to that subset is a
valid and effective means of data reduction. However, once our
interests extend beyond a single region or the region becomes
too large, resource constraints dictate that we cannot load the
entire data set into memory at once, and we need to iterate over the
data to reduce them to a set of interpretable summaries.

Iteration lends itself to parallelism, that is, computing on multiple
parts of the same problem simultaneously. Thus, in addition to
meeting memory constraints, iteration lets us leverage additional
processing resources to reduce overall computation time.
Investing in additional hardware is often more economical than
investment in software optimization. This is particularly relevant
in scientific computing, where we are faced with a diverse, rapidly
evolving set of unsolved problems, each requiring specialized
software. The costs of investment in general purpose hardware are
amortized over each problem, rather than paid each time for software
optimization. This also relates to maintainability: optimization
typically comes at a cost of increased code complexity. Many types
of summary and filter operations are cheap to implement in parallel
because the data partitions can be processed independently. We call
this type of operation \emph{embarrassingly} parallel. For example, the
counting of reads overlapping a gene does not depend on the counting
for a different gene.

Given the complexity and scope of the data, analysts often rely on
visual tools that display summaries and restricted views to
communicate information at different scales and level of detail,
from the whole genome to single nucleotide resolution. Plot
interactivity is always a useful feature when exploring data, and
this is particularly true with big data. The view is always
restricted in terms of its region and detail level, so, in order to
gain a broader and deeper understanding of the data, the viewer will
need to adjust the view, either by panning to a different region,
zooming to see more details or adjusting the parameters of the
summary step. The size of the genome and the range of scales make it
infeasible to pre-render every possible view. Thus, the views need
to be generated dynamically, in lazy reaction to the
user. Performance is an important factor in interpretability: slow
transitions distract the viewer and obfuscate relationships between
views. Dynamic generation requires fast underlying computations to
load, filter and summarize the data, and fast rendering to display
the processed data on the screen.

This paper describes strategies to surmount computational and
visualization challenges in the analysis of large genomic data and
how they have been implemented in the R programming language by a
number of packages from the Bioconductor project \cite{bioc}. We
will demonstrate their application to a real data set: the
whole-genome sequencing of the HapMap cell line NA12878, the
daughter in the CEU trio. The GATK project genotyped the sample
according to their best practices and included the calls in their
resource bundle, along with the alignments for chr20, one of the
shortest chromosomes. Realistically, one would analyze the data for
the entire genome, but the chr20 subset is still too large to be
processed on a commodity laptop and thus is sufficient for our
purposes.

\section{Limiting Resource Consumption}\label{sec-2}

Our ultimate goal is to process and summarize a large
data set in its entirety, and iteration enables this by limiting the resource
commitment at a given point in time. Limiting resource
consumption generalizes beyond iteration and is a fundamental
technique for computing with big data. In many cases, it may render
iteration unnecessary. Two effective approaches for being frugal
with data are restriction and compression. Restriction means
controlling which data are loaded and lets us avoid wasting
resources on irrelevant or excessive data. Compression helps by
representing the same data with fewer resources.

\subsection{Restricting Queries}\label{sec-2-1}

Restriction is appropriate when the question of interest requires
only a fraction of the available data. It is applicable in
different ways to sequence
vectors, range-based annotations and feature-by-sample matrices. We
can restrict data along two dimensions: row/record-wise and/or
column/attribute-wise, with genomic overlap being an important row
filter. Sequences and genomic vectors are relatively simple
structures that are often restricted by range, that is, extraction of
a contiguous subsequence of per-position values. Row-wise
restriction is useful when working with large sets of
experimentally generated short sequences. The sequence aligner
generates alignments as annotations on a reference sequence, and
these alignments have many attributes, such as genomic
position, score, gaps, sequence and sequence quality. Restriction
can exclude the irrelevant attributes. Analysts often slice large
matrices, such as those consisting of SNP calls, by both row (SNP)
and column (individual).

A special mode of restriction is to randomly generate a selection
of records. Down-sampling can address many questions, especially
during quality assessment and data exploration. For example, short
reads are initially summarized in FASTQ files containing a
plain text representation of base calls and corresponding quality
scores. Basic statistics of quality assessment such as the
nucleotide count as a function of sequencing cycle or overall GC
content are very well characterized by random samples of a million
reads, which might be 1\% of the data. This sample fits easily in
memory. Computations on this size of data are very nimble,
enabling interactive exploration on commodity computers. An
essential requirement is that the data represent a random sample.

The \emph{ShortRead} package is designed for the QA and exploratory
analysis of the output from high-througput sequencing
instruments. It defines the \texttt{FastqSampler} object, which draws
random samples from FASTQ files. The sequence reads in our data set
have been extracted into a FASTQ file from the publicly available
alignments. We wish to check a few quality statistics before
proceeding. We begin by loading a random sample of one million
reads from the file:\vspace*{6pt}\\
\mbox{
\includegraphics{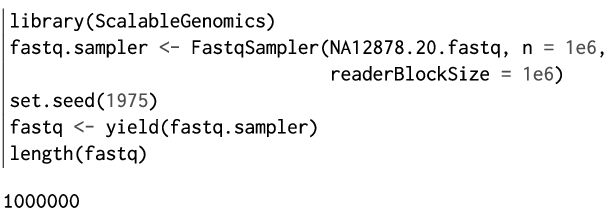}
}
%

With the sequences loaded, we can compute some QA statistics, like
the overall base call tally:\vspace*{6pt}\\
\mbox{
\includegraphics{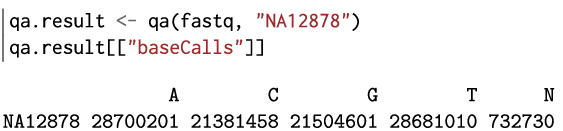}
}
%

In a complete workflow, we would generate an HTML QA report via the
\texttt{report} function.

An example of a situation where random sampling does \emph{not} work is
when prototyping a statistical method that depends on a significant
amount of data to achieve reasonable power. Variant calling is a
specific example: restricting the number of reads would lead to
less coverage, less power and less meaningful results. Instead, we
need to restrict the analysis to a particular region and include
all of the reads falling within it.

To optimize range-based queries, we often sort and index our data
structures by genomic coordinates. We should consider indexing an
investment because an index is generally expensive to generate but
cheap to query. The justification is that we will issue a sufficient
number of queries to outweigh the initial generation cost. Three
primary file formats follow this pattern: BAM, Tabix and BigWig
\cite{kent2010bigwig,li2009sequence}. Each format is best suited
for a particular type of data. The BAM format is specially designed
for sequence alignments and stores the complex alignment structure,
as well as the aligned sequence. Tabix is meant for indexing general
range-based annotations stored in tabular text files, such as BED
and GFF. Finally, BigWig is optimized for storing genome-length
vectors, such as the coverage from a sequencing experiment. BAM and
Tabix compress the primary data with block-wise gzip compression and
save the index as a separate file. BigWig files are similarly
compressed but are self-contained.

The \emph{Rsamtools} package is an interface between R and the
\emph{samtools} library, which implements access to BAM, Tabix and other
binary file formats. \emph{Rsamtools} enables restriction of BAM queries
through the \texttt{ScanBamParam} object. This object can be used as an
argument to all BAM input functions, and enables restriction to
particular fields of the BAM file, to specific genomic regions of
interest and to properties of the aligned reads (e.g., restricting
input to paired-end alignments that form proper pairs).

One common scenario in high-throughput sequencing is the
calculation of statistics such as coverage (the number of short
sequence reads overlapping each nucleotide in the genome). The data
required for this calculation usually come from very large BAM
files containing alignment coordinates (including the alignment
``cigar''), sequences and quality scores for tens of millions of
short reads. Only the smallest element of these data, the alignment
coordinates, is required for calculation of coverage. By
restricting input to alignment coordinates, we transform the
computational task from one of complicated memory management of
large data to simple vectorized operations on in-memory objects.

We can directly implement a coverage estimation by specifying a
\texttt{ScanBamParam} object that restricts to the alignment
information. The underlying coverage calculation is implemented by
the \emph{IRanges} package, which sits at the core of the Bioconductor
infrastructure and provides fundamental algorithms and data structures
for manipulating and annotating ranges. It is extended by
\emph{GenomicRanges} to add conveniences for manipulating ranges on the
genome:\vspace*{6pt}\\
\mbox{
\includegraphics{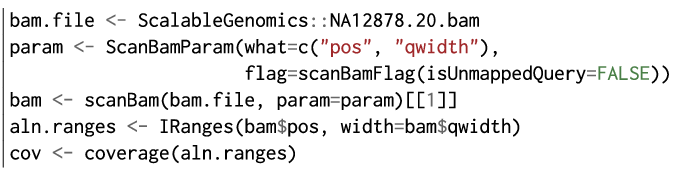}
}\\
This is only an estimate, however, because we have ignored the
complex structure of the alignments, for example, the insertions and
deletions. \emph{Rsamtools} provides a convenience function for the
more accurate calculation:\vspace*{6pt}\\
\mbox{
\includegraphics{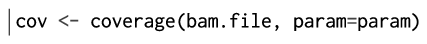}
}

\subsection{Compressing Genomic Vectors}\label{sec-2-2}

Some vectors, in particular, the coverage, have long stretches of
repeated values, often zeroes. An efficient compression scheme for
such cases is run-length encoding. Each run of repeated values is
reduced to two values: the length of the run and the repeated
value. This scheme saves space and also reduces computation time by
reducing computation size. For example, the vector ${0, 0, 0, 1, 1,
5, 5, 5}$ would have run-values ${0, 1, 5}$ and run-lengths ${3, 2,
3}$. The data have been reduced from a size of 8 to a size of 6 (3
values plus 3 lengths). The \emph{IRanges} \texttt{Rle} class is a run-length
encoded vector that supports the full R vector API on top of the
compressed representation. Operations on an \texttt{Rle} gain efficiency
by taking advantage of the compression. For example, the \texttt{sum}
method computes a sum of the run values, using the run lengths as
weights. Thus, the time complexity is on the order of the number of
runs, rather than the length of the vector.

The \texttt{cov} object we generated in the previous section is a list of
\texttt{Rle} objects, one per chromosome.\vspace*{6pt}\\
\mbox{
\includegraphics{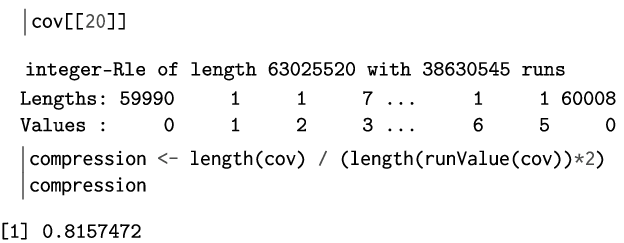}
}

For this whole-genome sequencing, the data are quite dense and
complex, so the compression actually decreases efficiency. However,
in the course of analysis we often end up with sparser data and
thus better compression ratios. In this analysis, we are concerned
about regions with extremely high coverage: these are often due to
alignment artifacts.\vspace*{6pt}\\
\mbox{
\includegraphics{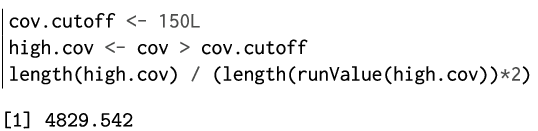}
}

Calculating the \texttt{sum} is then more efficient than with conventional
vectors:\vspace*{6pt}\\
\mbox{
\includegraphics{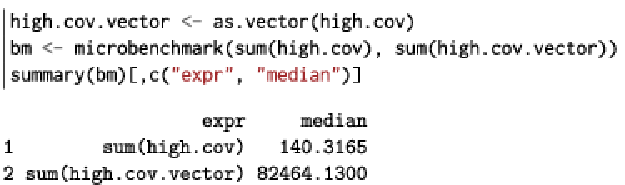}
}

Sometimes we are interested in the values of a genomic vector that
fall within a set of genomic features. Examples include the
coverage values within a set of called ChIP-seq peaks or the
conservation scores for a set of motif hits. We could extract the
subvectors of interest into a list. However, large lists bring
undesirable overhead, and the data would no longer be easily
indexed by genomic position. Instead,\vadjust{\goodbreak} we combine the original
vector with the ranges of interest. In \emph{IRanges}, this is called a
\texttt{Views} object. There is an \texttt{RleViews} object for
defining views on
top of an \texttt{Rle}.

To demonstrate, we \texttt{slice} our original coverage vector by our high
coverage cutoff to yield the regions of high coverage, overlaid on
the coverage itself, as an \texttt{RleViews} object:\vspace*{6pt}\\
\mbox{
\includegraphics{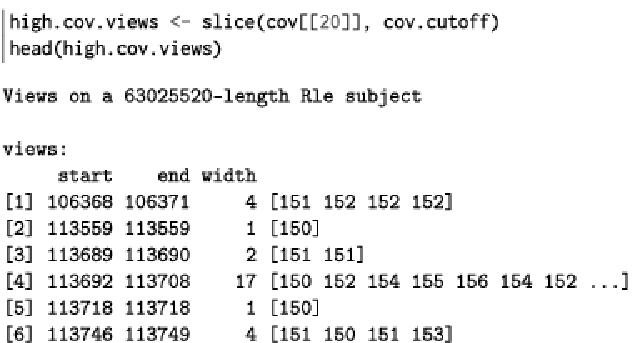}
}
%

This lets us efficiently calculate the average coverage in each
region:\vspace*{6pt}\\
\mbox{
\includegraphics{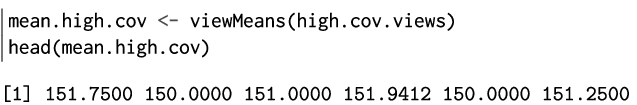}
}

The \emph{Biostrings} package \cite{Biostrings} provides \texttt{XString\-Views}
for views on top of DNA, RNA and amino acid sequences. \texttt{XString} is
a reference, rather than a value as is typical in R, so we can
create multiple \texttt{XStringViews} objects without copying the
underlying data. This is an application of the \emph{fly-weight} design
pattern: multiple objects decorate the same primary data structure,
which is stored only once in memory.

We can apply \texttt{XStringViews} for tabulating the nucleotides underlying
the high coverage regions. First, we need to load the sequence for
chr20, via the \emph{BSgenome} package and the addon package for human:\vspace*{6pt}\\
\mbox{
\includegraphics{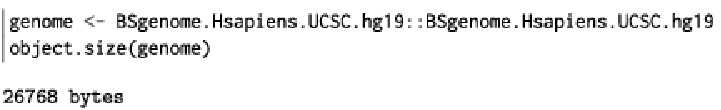}
}

While the human genome consists of billions of bases, our genome
object is tiny. This is an example of lazy loading: chromosomes are
loaded, and cached, as requested. In our case, we restrict to
chr20 and form the \texttt{XStringViews}.\vspace*{6pt}\\
\mbox{
\includegraphics{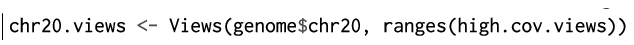}
}

We verify that the cached sequence occupies the same memory as
the subject of the views:\vspace*{6pt}\\
\mbox{
\includegraphics{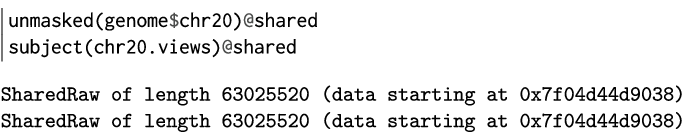}
}

Finally, we calculate and compare the nucleotide frequencies:\vspace*{6pt}\\
\mbox{
\includegraphics{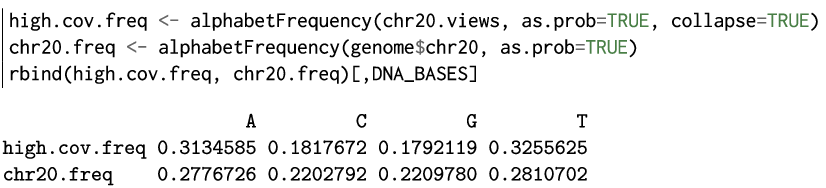}
}

We notice that the high coverage regions are A/T-rich, which
is characteristic of low complexity regions.

\subsection{Compressing Lists}\label{sec-2-3}

The high coverage regions in our data may be associated with the
presence of repetitive elements that confuse the aligner. We obtain
the repeat annotations from the UCSC genome browser with the
\emph{rtracklayer} package, which, in addition to a browser interface,
handles input and output for various annotation file formats,
including BigWig. Our query for the repeats is restricted to chr20,
which saves download time. We subset to the simple and low
complexity repeats, which are the most likely to be problematic:\vspace*{6pt}\\
\mbox{
\includegraphics{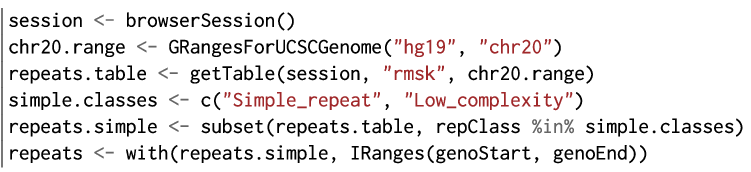}
}\\
Our goal is to calculate the percent of each high coverage region
covered by a repeat. First, we split the repeats according to
overlap with a high coverage region:\vspace*{6pt}\\
\mbox{
\includegraphics{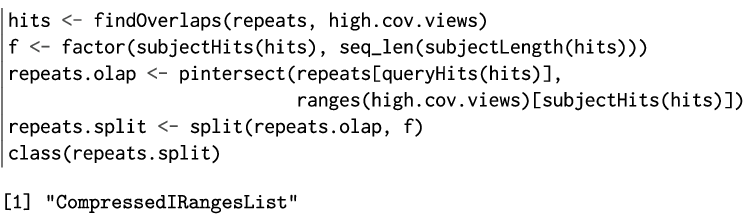}
}

The \texttt{repeats.split} object is not an ordinary list.

Long lists are expensive to construct, store and process. Creating a
new vector for each group requires time, and there is storage
overhead for each vector. Furthermore, data compression is less
efficient when the data are split across objects. Depending on the
implementation of the list elements, these costs can be
significant. This is particularly true of the S4 object system in R
\cite{chambers08}. Another detriment to R lists is that list
elements can be of mixed type. Thus, there are  few native
routines for computing on lists. For example, the R \texttt{sum} function
efficiently sums the elements of a homogeneous numeric vector, but
there is no support for calling \texttt{sum} to calculate the sum of each
numeric vector in a list. Even if such routines did exist for native
data types, there are custom data types, such as ranges, and we aim
to facilitate grouping of any data that we can model as a vector.

\begin{figure}

\includegraphics{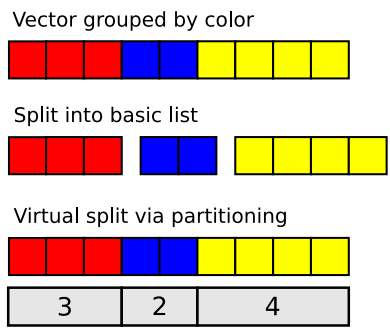}

\caption{Grouping via partitioning vs. splitting into multiple objects.
Top: the input vector, with elements belonging to three different
groups: red, blue and yellow. Middle: typical splitting of vector into
three vectors, one per group. This brings the overhead of multiple
objects. Bottom: the data are virtually split by a partitioning,
encoded by the number of elements in each group (the vector is assumed
to be sorted by group).}\label{fig:last-part}
\end{figure}

While the R \texttt{sum} function is incapable of computing group sums,
there is an oddly named function called \texttt{rowsum} that will
efficiently compute them, given a numeric vector and a grouping
factor. This hints that a more efficient approach to grouping may
be to store the original vector along with a partitioning. The
\emph{IRanges} R package includes a \texttt{CompressedList} framework that
follows this strategy. A \texttt{CompressedList} consists of the data
vector, sorted by group, and a vector of indexes where each group
ends in the data vector (see Figure~\ref{fig:last-part}). \emph{IRanges}
provides \texttt{CompressedList} implementations for native atomic vectors
and other data types in the \emph{IRanges} infrastructure, and the
framework is extensible to new data types. A \texttt{CompressedList} is
homogeneous, so it is natural to define methods on subclasses to
perform operations particular to a type of data. For example, there
is a \texttt{sum} method for the \texttt{NumericList} class that delegates
internally to \texttt{rowsum}. This approach bears similarity to storing
data by columns: we improve storage efficiency by storing fewer
objects, and we maintain the data in its most readily computable
form. It is also an application of \emph{lazy} computing, where we delay
the partitioning of the data until a computation requires it. We
are then in position to optimize the partitioning according to the
specific requirements of the operation.

Since \texttt{repeats.list} is a \texttt{CompressedList}, we can take
advantage of
these optimizations:\vspace*{6pt}\\
\mbox{
\includegraphics{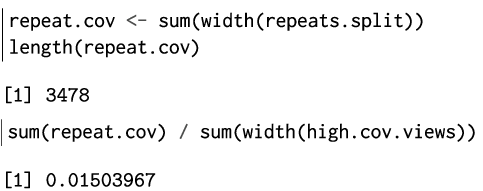}
}

This value can be compared to the percent of chr20 covered:\vspace*{6pt}\\
\mbox{
\includegraphics{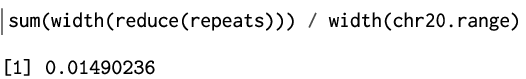}
}

Instead of a \texttt{CompressedList}, we could have {\spaceskip=0.2em plus 0.05em minus 0.04em solved this problem
using \texttt{coverage} and \texttt{RleViews}.}

The downside of compression is that there is overhead to explicit
iteration because we need to extract a new vector with each step.
The \emph{Biostrings} package has explored a solution. We can convert our
\texttt{XStringViews} object \texttt{chr20.views} to a \texttt
{DNA\-StringSet} that
contains one \texttt{DNAString} for each view window. The data for each
\texttt{DNAString} has never been copied from the original chr20 sequence,
and any operations on a \texttt{DNAString} operate directly on the shared
data. While this solution may seem obvious, it relies heavily on
native code and is far from the typical behavior of R data structures:\vspace*{6pt}\\
\mbox{
\includegraphics{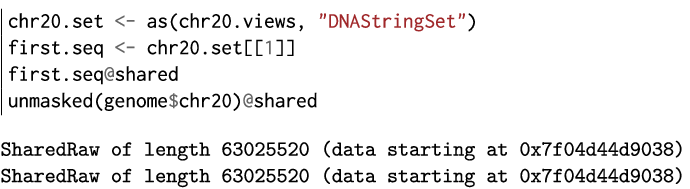}
}

The nascent \emph{XVector} package aims to do the same for other R data
types, such as integer, double and logical values.

\section{Iterating}\label{sec-3}
\subsection{Splitting Data}\label{sec-3-1}

Iterative summarization of data may be modeled as three separate
steps: split, apply and combine
\cite{Wickham:2010:JSSOBK:v40i01}. The split step is typically the
only one that depends on the size of the input data. The apply step
operates on data of restricted size, and it should reduce the data
to a scale that facilitates combination. Thus, the most challenging
step is the first: splitting the data into chunks small enough to
meet resource constraints.

Two modes of splitting are particularly applicable to genomic data:
sequential chunking and genomic partitioning. Sequential chunking
is a popular and general technique that simply loads records in
fixed-count chunks, according to the order in which they are
stored. Genomic partitioning iterates over a disjoint set of ranges
that cover the genome. Typical partitioning schemes include one
range per chromosome and sub-chromosomal ranges of some uniform
size. Efficient range-based iteration, whether over a partitioning
or list of interesting regions, depends on data structures, file
formats and algorithms that are optimized for range-based queries.

Under the assumption that repeat regions are leading to anomalous
alignments, we aim to filter from our BAM file any alignments
overlapping a repeat. As we will be performing many overlap queries
against the repeat data set, it is worth indexing it for faster
queries. The algorithms for accessing BAM, Tabix and BigWig files
are designed for genome browsers and have not been optimized for
processing multiple queries in a batch. Each query results in a new
search. This is unnecessarily wasteful, at least when the query
ranges are sorted, as is often the case. We could improve the
algorithm by detecting whether the next range is in the same bin
and, if so, continuing the search from the current position. The
\emph{IRanges} package identifies interval trees \cite{Cormen2001} as an
appropriate and well-understood data structure for range-based
queries, and implements these using a combination of existing C
libraries \cite{kent2002human} and new C source code. The query is
sorted, and every new search begins at the current node, rather
than at the root. We build a \texttt{GIntervalTree} for the repeats:\vspace*{6pt}\\
\mbox{
\includegraphics{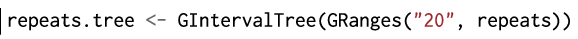}
}\\
The \texttt{GIntervalTree} from \emph{GenomicRanges} enables the same
optimization when data are aligned to multiple chromosomes.

To configure streaming, we specify a \texttt{yieldSize} when constructing
the object representing our BAM file. We will filter at the
individual read level, but it should be noted that for paired-end
data \emph{Rsamtools} supports streaming by read pair, such that members
of the same pair are guaranteed to be in the same chunk. Since BAM
files are typically sorted by position, not pair, this is a
significant benefit:\vspace*{6pt}\\
\mbox{
\includegraphics{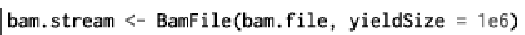}
}

To filter the BAM, we first need to define a filter rule that
excludes reads that overlap a repeat. The low-level \emph{Rsamtools}
interface provides the read data as a \texttt{DataFrame}, which we convert
into a \texttt{GAlignments} object from the \emph{GenomicAlignments} package,
which provides data structures and utilities for analyzing
read alignments:\vspace*{6pt}\\
\mbox{
\includegraphics{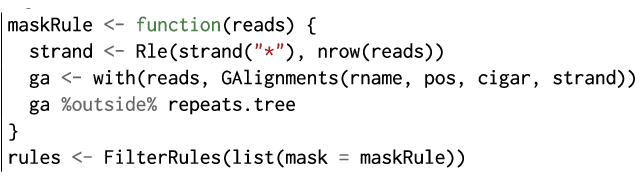}
}\\
Since we are writing a new BAM file, this is iteration with a
side effect rather than a reduction.

To demonstrate reduction, we will calculate the coverage in an
iterative fashion, which ends up identical to our original
calculation:\vspace*{6pt}\\
\mbox{
\includegraphics{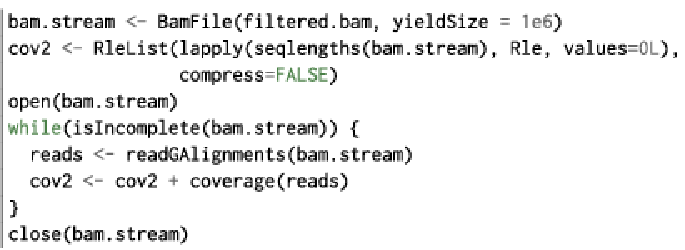}
}

Choosing an appropriate yield size for each iteration is
important. There is overhead to each iteration, mostly due to I/O
and memory allocation, as well as the R evaluator. Thus, one
strategy is to increase the size of each iteration (and reduce the
number of iterations) until the data fit comfortably in memory. It
is relatively easy to estimate a workable yield size from the
consumption of processing a single chunk. The \texttt{gc} function
exposes the
maximum amount of memory consumed by R between resets:\vspace*{6pt}\\
\mbox{
\includegraphics{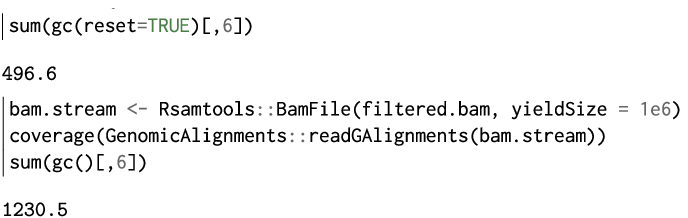}
}

The memory usage started at about 500 MB and peaked at about 1200
MB, so the iteration consumed up to 700 MB. With 8~GB of ram, we
might be able to process up to 10 million reads at once, assuming
linear scaling.

As an alternative to streaming over chunks, we can iterate over a
partitioning of the genome or other domain. Genomic partitioning
can be preferable to streaming when we are only interested in
certain regions. The \texttt{tileGenome} function is a convenience for
generating a set of ranges that partition a genome. We rely on it to
reimplement the \texttt{coverage} iterative calculation with a
partitioning:\vspace*{6pt}\\
\mbox{
\includegraphics{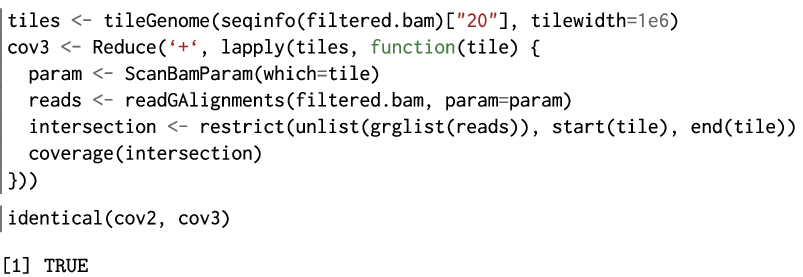}
}

A caveat with partitioning is that since many query algorithms
return ranges with any overlap of the query, care must be taken to
intersect the results with each partition, so that reads are
not double counted, for example.

By computing the coverage, we have summarized the data. Computing
summaries is often time consuming, but since the summaries are smaller than
the original data, it is feasible to store them for later
use. Caching the results of computations is an optimization
technique known as \emph{memoization}. An analysis rarely follows a
linear path. By caching the data at each stage of the analysis, as
we proceed from the raw data to a feature-level summary, often
with multiple rounds of feature annotation, we can avoid redundant
computation when we inevitably backtrack and form branches. This
is an application of \emph{incremental computing}.
We export our coverage as a BigWig file, for later use:\vspace*{6pt}\\
\mbox{
\includegraphics{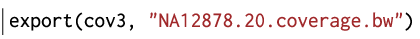}
}

\subsection{Iterating in Parallel}\label{sec-3-2}

There are two basic modes of parallelism: data-level and
task-level. Embarrassingly parallel problems illustrate data
parallelism. Work flows might less frequently involve task
parallelism, where different tasks are applied to the same data
chunk. These are generally more challenging to implement,
especially with R, which does not offer any special support for
concurrency. The \emph{Streamer} package has explored this direction.

Multicore and cluster computing are similar in that they are
modular, and scaling algorithms to use multiple cores or multiple
nodes can involve conceptually similar steps, but there are some
critical differences. Multiple cores in the same system share the
same memory, as well as other resources. Shared memory
configurations offer fast inter-thread data transfer and
communication. However, the shared resources can quickly become
exhausted and present a bottleneck. Computing on a cluster
involves significant additional expertise to access and manage
cluster resources that are shared between multiple users and
governed by a scheduler. Interacting with a scheduler introduces an
extra step into a workflow. We place jobs in a queue, and the jobs
are executed asynchronously. Another complication is that we need
to share the data between every computer. A naive but often
sufficient method is to store the data in a central location on a
network file system and to distribute the data via the
network. The network overhead implied by this approach may penalize performance.

When the ratio of communication to computation time is large,
communication dominates the overall calculation. The main
strategies for addressing this are to (a) ensure each task
represents a significant amount of work and (b) identify points
where data sizes of inputs (e.g., file names) and outputs (e.g.,
vector of counts across regions of interest) are small. Data
partitioning is usually conveyed to workers indirectly, for example, via
specification of the range of data to be processed, rather than
inputting and explicitly partitioning data. This approach reduces
the communication costs between the serial and parallel portions of
the computation and avoids loading the entire data set into memory.

The R packages \emph{foreach} \cite{foreach}, \emph{parallel} (distributed
with R \cite{rman}), \emph{pbdR} \cite{pbdR2012} and \emph{BatchJobs}
\cite{BatchJobs} provide
abstractions and implementations for executing tasks in parallel
and support both the shared memory and cluster
configurations. BatchJobs and \emph{pbdR} are primarily designed for
asynchronous
execution, where jobs are submitted to a scheduling system, and the
user issues commands to query for job status and collect results
upon completion. The other two, \emph{foreach} and \emph{parallel},
follow a
synchronous model conducive to interactive use.

Different use cases and hardware configurations benefit from
different parallelization strategies. An analyst might apply
multiple strategies in the course of an analysis. This has
motivated the development of an abstraction oriented toward
genomics workflows. The \emph{BiocParallel} package defines this
abstraction and implements it on top of \emph{BatchJobs}, \emph
{parallel} and
\emph{foreach} to support the most common configurations. An important
feature of \emph{BiocParallel} is that it encapsulates the
parallelization strategy in a parameter object that can be passed
down the stack to infrastructure routines that implement the
iteration. Thus, for common use cases the user can take advantage
of parallelism by solely indicating the appropriate
implementation. Iteration is carried out in a functional manner, so
the API mirrors the *apply functions in base R: \texttt{bplapply},
\texttt{bpmapply}, etc.

To illustrate use of parallel iteration, we diagnose the GATK genotype
calls introduced earlier. One approach is to generate our own set of nucleotide
tallies, perform some simple filtering to yield a set of variant
calls, and compare our findings to those from GATK. The set of
nucleotide tallies is a more detailed form of the coverage that
consists of the count of each nucleotide at each position, as well
as some other per-position statistics. Tallies are useful for
detecting genetic variants through comparison to a reference
sequence.

The \emph{VariantTools} package provides a facility for summarizing the
nucleotide counts from a BAM file over a given range. We can
iterate over the tiling in parallel using the \texttt{bplapply}
function. The \texttt{BPPARAM} argument specifies the parallel
implementation. \texttt{MulticoreParam} is appropriate for a multicore
workstation, whereas we might use \texttt{BatchJobsParam} for scheduling
each iteration as a job on a cluster:\vspace*{6pt}\\
\mbox{
\includegraphics{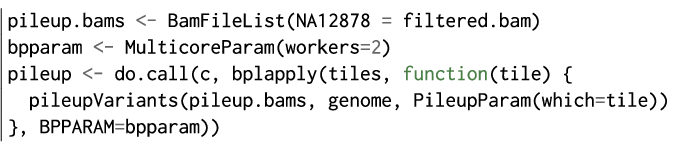}
}\\
The above is an example of explicit iteration. Thanks to the
encapsulation and abstraction afforded by \texttt{BiocParallelParam}, the
\texttt{pileupVariants} function supports parallel iteration directly, so
the implementation becomes much simpler:\vspace*{6pt}\\
\mbox{
\includegraphics{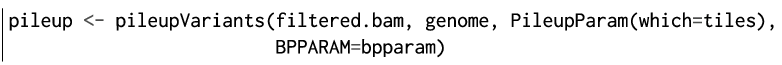}
}\\
This is an example of an embarrassingly parallel solution: each
iteration is a simple counting exercise and is independent of the
others. An example of a \emph{non}embarrassingly parallel algorithm is
our demonstration of BAM filtering: each iteration has the
side effect of writing to the same file on disk. The increased
complexity of coordinating the I/O across jobs undermines the value
of parallelism in that case.
\section{Scaling Genomic Graphics}
\label{sec-4}

\subsection{Managing Graphical Resources}\label{sec-4-1}

Graphics software is special in that it performs two roles:
distilling information from the data and visually communicating
that information to the user. The first role is similar to any data
processing pipeline; the unique aspect is the communication. The
communication bandwidth of a plot is limited by the size and
resolution of the display device and the perceptive capabilities of
the user. These limitations become particularly acute in genomics,
where it would be virtually impossible to communicate the details
of a billion alignments along a genome of 3 billion nucleotides.

When considering how best to manage graphical resources, we recall
the general technique of restriction. Restriction has obvious
applicability to genomic graphics: we can balance the size of the
view and the level of detail. As we increase the size of the view,
we must decrease the level of detail and vice versa. This means
only so much information can be communicated in a single plot, so
the user needs to view many plots in order to comprehend the
data. It would be infeasible to iteratively generate every possible
plot, so we need to lazily generate plots in response to user
interaction. For example, the typical genome browser supports
panning and zooming about the genome, displaying data at different
levels of detail, depending on the size of the genomic region.

\begin{figure*}

\includegraphics{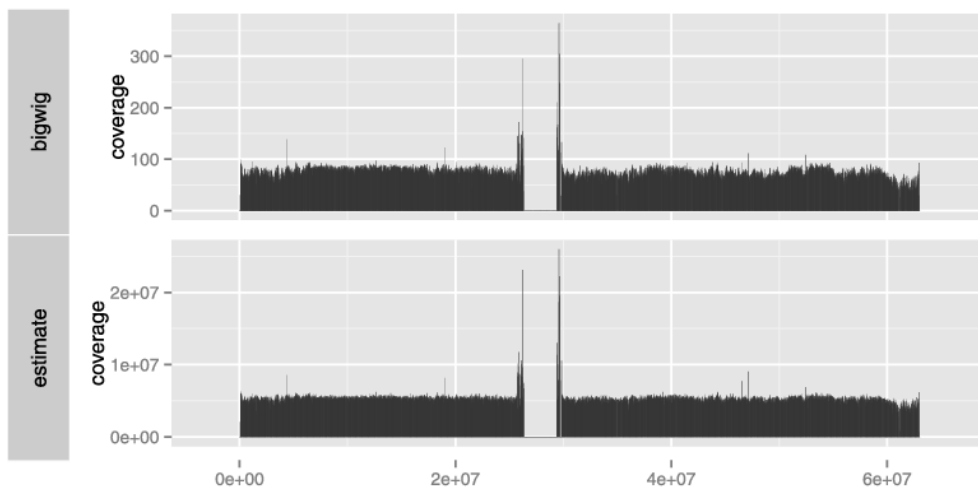}

\caption{The results of two coverage calculations over chr20. Top: the
calculation based on cached values in the BigWig file. Bottom: the
estimated coverage from the BAM index file.}\label{fig:cov}
\end{figure*}

\subsection{Displaying Summaries Efficiently}\label{sec-4-2}

When plotting data along a restricted range, graphics software can
rely on the support for range-based queries presented
previously. Controlling the level of detail is more challenging
because it relies on summaries. As the viewed region can be as
large as the genome, generating summaries is often computationally
intensive and introduces undesirable latency to plot updates. One
solution to this problem is caching summaries at different levels
of detail. Global summaries will be regularly accessed and
expensive to compute, and thus are worth caching, whereas the
detailed data exposed upon drill-down can be computed lazily. This
strategy is supported by the BigWig format. In addition to storing
a full genomic vector, BigWig files also contain summary vectors,
computed over a range of resolutions, according to the following
statistics: mean, min, max and standard deviation.\vadjust{\goodbreak} Plotting the
aggregate coverage is a shortcut that avoids pointless rendering of
data that is beyond the display resolution and the perceptive
abilities of the viewer:\vspace*{6pt}\\
\mbox{
\includegraphics{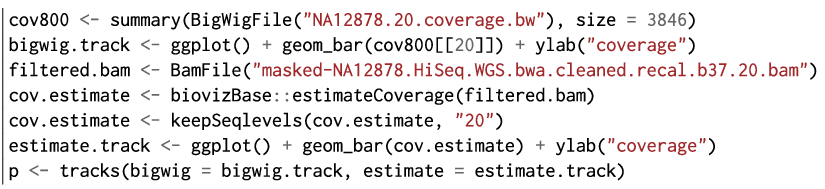}
}


A good summary will guide the user to the most interesting parts of
the data. Genomic data are typically sparsely distributed along the
genome, due to the nonuniform distribution of genes and
experimental protocols that enrich for regions of
interest. Coverage is a particularly useful summary, as it helps
guide the viewer to the regions with the most data. The following
gets the average coverage for 800 windows (perhaps appropriate for
an 800 pixel plot). The result is shown in the top panel of
Figure~\ref{fig:cov}:\vspace*{6pt}\\
\mbox{
\includegraphics{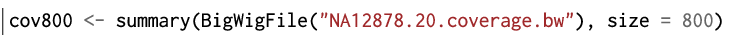}
}

In cases where a BigWig file or other cached summary is
unavailable, we can rely on a heuristic that estimates the coverage
from the index of a BAM or Tabix file. The index stores offsets
into the BAM for efficient range-based queries. Instead of
accessing the index to resolve queries, we calculate the difference
in the file offsets for each range and derive a relative coverage
estimate at a coarse level of resolution. In practice, this reduces
the required time to compute the coverage from many minutes to a
few seconds. When the plot resolution exceeds the resolution of the
index, we again rely on the index to query the BAM file for the
reads that fall within the relatively small region and compute the
coverage directly. A heuristic seems acceptable in this case,
because improved accuracy is immediately accessible by
zooming. This is in contrast to pure statistical computations,
where crude estimates are less appreciated, even in the exploratory
context, since resolution is not so readily forthcoming.

The \texttt{estimateCoverage} function from the \emph{bio\-vizBase} package
\cite{biovizBase} estimates the coverage from the BAM index
file. The bottom panel of Figure~\ref{fig:cov} shows the output of
\texttt{estimateCoverage} for the example data set and allows for
comparison with the more exact calculation derived from the BigWig
file. The two results are quite similar and both required only a
few seconds to compute on a commodity laptop.
\subsection{Generating Plots Dynamically}
\label{sec-4-3}

\begin{figure*}[b]

\includegraphics{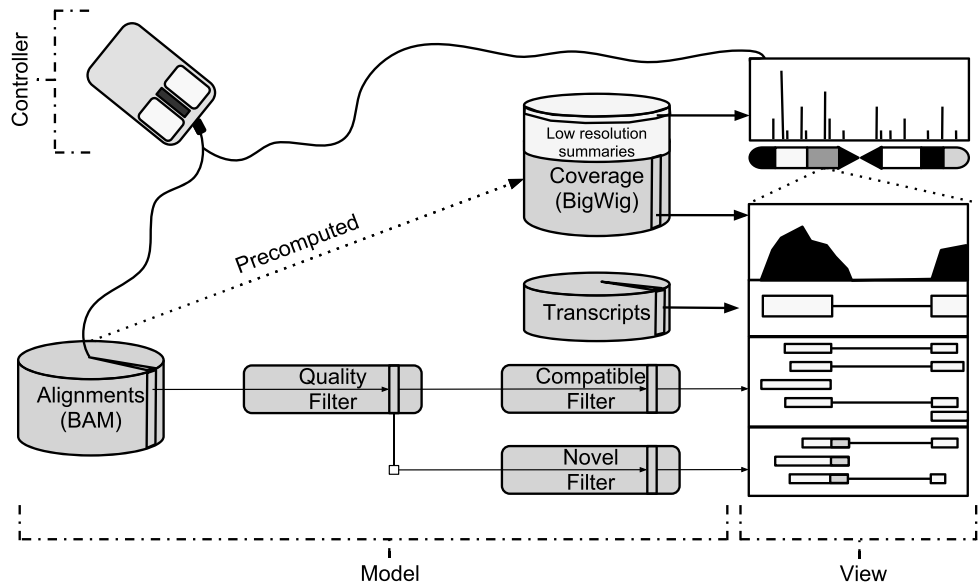}

\caption{An application of the model--view--controller pattern and
pre-computed summaries to genomic visualization. Coverage is displayed
at two levels of resolution (whole chromosome and the current zoom)
after efficient extraction from the multi-resolution BigWig file. The
BAM file holding the read alignments is abstracted by a multi-stage
data model, consisting of the BAM source, a dynamic read quality filter
and two filters that effectively split the alignments according to
compatibility with the known transcript annotations. The view contains
several coordinately-zoomed plots, as well as an ideogram and coverage
overview. Each plot obtains its data from one of the data model
components. The controller might adjust the data model filter settings
and the current zoom in response to user commands.}\label{fig:MVC-genomics}
\end{figure*}

\begin{figure*}

\includegraphics[scale=0.99]{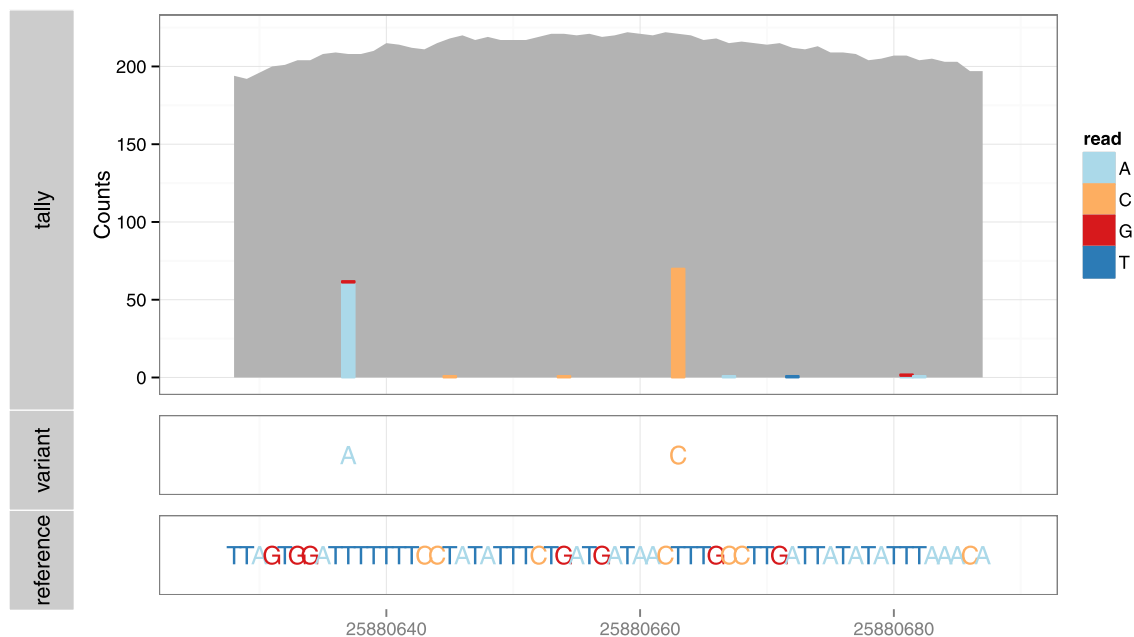}

\caption{Example plot for diagnosing genotype calls, consisting of the
nucleotide tallies, genotype calls and reference sequence, from top to
bottom. The plot is dynamically generated for the selected region of
interest, without processing the entire genome. The viewer might check
to see if the tallies support the called genotypes. In this case, the
data are consistent.}\label{fig:tracks}
\end{figure*}

The design of interactive graphics software typically follows the
model--view--controller pattern (see Figure~\ref{fig:MVC-genomics}).
The\vadjust{\goodbreak}
view renders data retrieved from the data model, and the controller
is the interface through which the user manipulates the view and
data model. The data model abstracts the underlying data source,
which might be memory, disk or a dynamic computation. The
abstraction supports the implementation of complex optimizations
without exposing any of the complexity to client code. Data is
communicated to the user through the view, and user input is
received through the controller. A complex application will
consist of multiple interactive views, linked through a common data
model, itself composed of multiple modules, chained together as
stages in a pipeline. The viewer, plots and pipeline stages are
interlinked to form a network.

A simple data model abstracts access to the primary data, such as an
in-memory GRanges object of transcript annotations or a BAM file
on disk. We can extend the simple model to one that dynamically
computes on data as they are requested by the application. This is
an example of lazy computing. Each operation is encapsulated into a
data model that proxies an underlying model. The proxy models form
a chain, leading from the raw data to the processed data that are
ready for plotting \cite{wickham2009plumbing}. Dynamic computation
avoids unnecessarily processing the entire genome when the user is
only interested in a few small regions, especially when the
parameters of the transformations frequently change during the
session. The data may be cached as they are computed, and the
pipeline might also anticipate future requests; for example, it
might prepare the data on either side of the currently viewed
region, in anticipation of scrolling. Caching and prediction are
examples of complex optimizations that are hidden by the data
model. The \emph{plumbr} R package \cite{plumbr} provides a proxy model
framework for implementing these types of ideas behind the data
frame API.

We have been experimenting with extending these approaches to
genomic data. The \emph{biovizBase} package implements a graphics-friendly
API for restricted queries to Bioconductor-supported data sources.
The \emph{ggbio} package builds on \emph{biovizBase} to support genomic plot
objects that are regenerated as the user adjusts the viewport.

To diagnose the GATK genotype calls, we combine the reference
sequence, nucleotide pileup and the genotype calls. The result is
shown in Figure~\ref{fig:tracks}. The \emph{ggbio} package produced the
plot by relying on restricted query support in \emph{biovizBase}. We
have already introduced the extraction of\vadjust{\goodbreak} genomic sequence and the
calculation of nucleotide pileups. The genotypes were drawn by the
\emph{VariantAnnotation} package from a Variant Call Format (VCF,
\cite{danecek2011variant}) file with a range-based index provided
by Tabix.

To generate the plot, we first select the region of interest:\vspace*{6pt}\\
\mbox{
\includegraphics{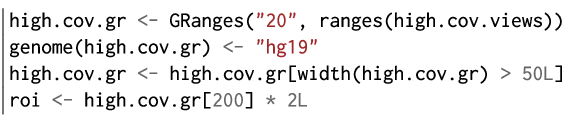}
}

Next, we construct the plot object and render it:\vspace*{6pt}\\
\mbox{
\includegraphics{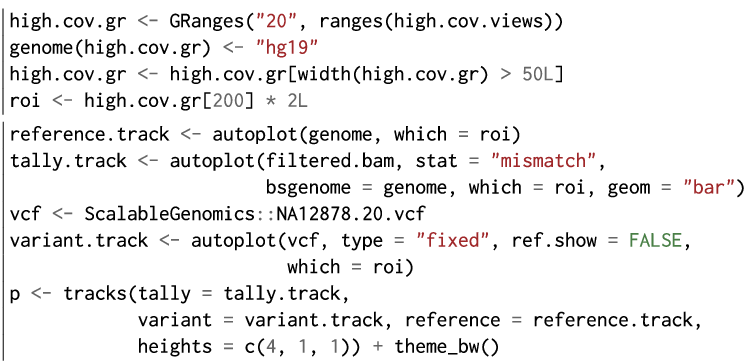}
}\\
Since the plot object is a logical representation of the plot,
that is, it references the original data, we can adjust various
aspects of it and generate a new rendering. In particular, we can
change the currently viewed region, and the data for the new region
are processed dynamically to generate the new plot. In this
example, we zoom out to a larger region around the first region:\vspace*{6pt}\\
\mbox{
\includegraphics{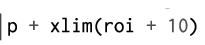}
}

Current work is focused on the \emph{MutableRanges} package, which
generalizes and formalizes the designs in
\emph{biovizBase}. It defines dynamic versions of the \emph{GenomicRanges}
data structures, for example, there is a \emph{DynamicGRanges} that
implements the \emph{GRanges} API on top of a BAM file. Only the
requested regions are loaded, and they are optionally cached for
future queries. A \emph{ProxyGRanges} performs dynamic computations based
on another \emph{GRanges}. This will enable a new generation of
interactive genomic visualization tools in R. An early adopter is
\emph{epivizr}, the R interface to the web-based \emph{epiviz}, a web-based
genome browser with support for general interactive graphics,
including scatterplots and histograms.

\section{Conclusion}\label{sec-5}
We have introduced software and techniques for \mbox{analyzing} and
plotting big genomic data. The Bioconductor project distributes the
software as a number of different R packages, including \emph{Rsamtools},
\emph{IRanges}, \emph{GenomicRanges}, \emph{GenomicAlignments},\break  \emph
{Biostrings},
\emph{rtracklayer}, \emph{biovizBase} and \emph{BiocParallel}. The
software enables
the analyst to conserve computational resources, iteratively
generate summaries and visualize data at arbitrary levels of
detail. These advances have helped to ensure that R and Bioconductor
remain relevant in the age of high-throughput sequencing. We plan to
continue in this direction by designing and implementing
abstractions that enable user code to be agnostic to the mode of
data storage, whether it be memory, files or databases. This will
bring much needed agility to resource allocation and will enable the
user to be more resourceful, without the burden of increased
complexity.

\section*{Acknowledgments}
Supported in part by the National Human
Genome Research Institute of the National Institutes of Health
(U41HG004059 to M. M.) and the National Science
Foundation (1247813 to M. M.).


%


\begin{thebibliography}{17}

\bibitem{BatchJobs}
%
\begin{bmisc}[auto:STB|2014/05/27|07:42:02]
\bauthor{\bsnm{Bischl},~\bfnm{Bernd}\binits{B.}},
\bauthor{\bsnm{Lang},~\bfnm{Michel}\binits{M.}},
\bauthor{\bsnm{Mersmann},~\bfnm{Olaf}\binits{O.}},
\bauthor{\bsnm{Rahnenfuehrer},~\bfnm{Joerg}\binits{J.}} \AND
\bauthor{\bsnm{Weihs},~\bfnm{Claus}\binits{C.}}
(\byear{2011}).
\bhowpublished{Computing on high performance clusters with R: Packages
BatchJobs and BatchExperiments. Technical Report 1, TU Dortmund}.
\end{bmisc}
%
\bptok{imsref}%
\endbibitem

\bibitem{chambers08}
%
\begin{bbook}[auto:STB|2014/05/27|07:42:02]
\bauthor{\bsnm{Chambers},~\bfnm{John~M.}\binits{J.~M.}}
(\byear{2008}).
\btitle{Software for Data Analysis: Programming with R}.
\bpublisher{Springer},
\blocation{New York}.
\end{bbook}
%
\bptok{imsref}%
\endbibitem

\bibitem{Cormen2001}
%
\begin{bbook}[mr]
\bauthor{\bsnm{Cormen},~\bfnm{Thomas~H.}\binits{T.~H.}},
\bauthor{\bsnm{Leiserson},~\bfnm{Charles~E.}\binits{C.~E.}},
\bauthor{\bsnm{Rivest},~\bfnm{Ronald~L.}\binits{R.~L.}} \AND
\bauthor{\bsnm{Stein},~\bfnm{Clifford}\binits{C.}}
(\byear{2001}).
\btitle{Introduction to Algorithms},
\bedition{2nd} ed.
\bpublisher{McGraw-Hill, Boston, MA}.
\bid{mr={1848805}}
\end{bbook}
%
\bptok{imsref}%
\endbibitem

\bibitem{danecek2011variant}
%
\begin{barticle}[pbm]
\bauthor{\bsnm{Danecek},~\bfnm{Petr}\binits{P.}},
\bauthor{\bsnm{Auton},~\bfnm{Adam}\binits{A.}},
\bauthor{\bsnm{Abecasis},~\bfnm{Goncalo}\binits{G.}},
\bauthor{\bsnm{Albers},~\bfnm{Cornelis~A.}\binits{C.~A.}},
\bauthor{\bsnm{Banks},~\bfnm{Eric}\binits{E.}},
\bauthor{\bsnm{DePristo},~\bfnm{Mark~A.}\binits{M.~A.}},
\bauthor{\bsnm{Handsaker},~\bfnm{Robert~E.}\binits{R.~E.}},
\bauthor{\bsnm{Lunter},~\bfnm{Gerton}\binits{G.}},
\bauthor{\bsnm{Marth},~\bfnm{Gabor~T.}\binits{G.~T.}},
\bauthor{\bsnm{Sherry},~\bfnm{Stephen~T.}\binits{S.~T.}},
\bauthor{\bsnm{McVean},~\bfnm{Gilean}\binits{G.}},
\bauthor{\bsnm{Durbin},~\bfnm{Richard}\binits{R.}} \AND
\bauthor{\bsnm{{1000 Genomes Project Analysis Group}}}
(\byear{2011}).
\btitle{The variant call format and VCFtools}.
\bjournal{Bioinformatics}
\bvolume{27}
\bpages{2156--2158}.
\bid{doi={10.1093/bioinformatics/btr330}, issn={1367-4811},
pii={btr330}, pmcid={3137218}, pmid={21653522}}
\end{barticle}
%
\bptok{imsref}%
\endbibitem

\bibitem{bioc}
%
\begin{barticle}[auto:STB|2014/05/27|07:42:02]
\bauthor{\bsnm{Gentleman},~\bfnm{Robert~C.}\binits{R.~C.}},
\bauthor{\bsnm{Carey},~\bfnm{Vincent~J.}\binits{V.~J.}},
\bauthor{\bsnm{Bates},~\bfnm{Douglas~M.}\binits{D.~M.}} \AND
\bauthor{\bsnm{others}}
(\byear{2004}).
\btitle{Bioconductor: Open software development for computational
biology and bioinformatics}.
\bjournal{Genome Biol.}
\bvolume{5}
\bpages{R80}.
\end{barticle}
%
\bptok{imsref}%
\endbibitem

\bibitem{kent2002human}
%
\begin{barticle}[pbm]
\bauthor{\bsnm{Kent},~\bfnm{W.~James}\binits{W.~J.}},
\bauthor{\bsnm{Sugnet},~\bfnm{Charles~W.}\binits{C.~W.}},
\bauthor{\bsnm{Furey},~\bfnm{Terrence~S.}\binits{T.~S.}},
\bauthor{\bsnm{Roskin}, \bfnm{Krishna~M.}\binits{K.~M.}},
\bauthor{\bsnm{Pringle},~\bfnm{Tom~H.}\binits{T.~H.}},
\bauthor{\bsnm{Zahler},~\bfnm{Alan~M.}\binits{A.~M.}} \AND
\bauthor{\bsnm{Haussler},~\bfnm{David}\binits{D.}}
(\byear{2002}).
\btitle{The human genome browser at UCSC}.
\bjournal{Genome Res.}
\bvolume{12}
\bpages{996--1006}.
\bid{doi={10.1101/gr.229102. Article published online before print in
May 2002}, issn={1088-9051}, pmcid={186604}, pmid={12045153}}
\end{barticle}
%
\bptok{imsref}%
\endbibitem

\bibitem{kent2010bigwig}
%
\begin{barticle}[auto:STB|2014/05/27|07:42:02]
\bauthor{\bsnm{Kent},~\bfnm{W.~J.}\binits{W.~J.}},
\bauthor{\bsnm{Zweig},~\bfnm{A.~S.}\binits{A.~S.}},
\bauthor{\bsnm{Barber},~\bfnm{G.}\binits{G.}},
\bauthor{\bsnm{Hinrichs},~\bfnm{A.~S.}\binits{A.~S.}} \AND
\bauthor{\bsnm{Karolchik},~\bfnm{D.}\binits{D.}}
(\byear{2010}).
\btitle{BigWig and BigBed: Enabling browsing of large distributed datasets}.
\bjournal{Bioinformatics}
\bvolume{26}
\bpages{2204--2207}.
\end{barticle}
%
\bptok{imsref}%
\endbibitem

\bibitem{pmid23950696}
%
\begin{bmisc}[auto:STB|2014/05/27|07:42:02]
\bauthor{\bsnm{Lawrence},~\bfnm{Michael}\binits{M.}},
\bauthor{\bsnm{Huber},~\bfnm{Wolfgang}\binits{W.}},
\bauthor{\bsnm{Pag{\`e}s},~\bfnm{Herv{\'e}}\binits{H.}},
\bauthor{\bsnm{Aboyoun},~\bfnm{Patrick}\binits{P.}},
\bauthor{\bsnm{Carlson},~\bfnm{Marc}\binits{M.}},
\bauthor{\bsnm{Gentleman},~\bfnm{Robert}\binits{R.}},
\bauthor{\bsnm{Morgan},~\bfnm{Martin}\binits{M.}} \AND
\bauthor{\bsnm{Carey}, \bfnm{Vincent}\binits{V.}}
(\byear{2013}).
\bhowpublished{Software for computing and annotating genomic ranges.
\textit{PLoS Computational Biology}
\textbf{9}
{e1003118}}.
\end{bmisc}
%
\bptok{imsref}%
\endbibitem

\bibitem{plumbr}
%
\begin{bmisc}[auto:STB|2014/05/27|07:42:02]
\bauthor{\bsnm{Lawrence},~\bfnm{Michael}\binits{M.}} \AND
\bauthor{\bsnm{Wickham},~\bfnm{Hadley}\binits{H.}}
(\byear{2012}).
\bhowpublished{plumbr: Mutable and dynamic data models. R package
version 0.6.6.}
\end{bmisc}
%
\bptok{imsref}%
\endbibitem

\bibitem{li2009sequence}
%
\begin{barticle}[pbm]
\bauthor{\bsnm{Li},~\bfnm{Heng}\binits{H.}},
\bauthor{\bsnm{Handsaker},~\bfnm{Bob}\binits{B.}},
\bauthor{\bsnm{Wysoker},~\bfnm{Alec}\binits{A.}},
\bauthor{\bsnm{Fennell},~\bfnm{Tim}\binits{T.}},
\bauthor{\bsnm{Ruan},~\bfnm{Jue}\binits{J.}},
\bauthor{\bsnm{Homer},~\bfnm{Nils}\binits{N.}},
\bauthor{\bsnm{Marth},~\bfnm{Gabor}\binits{G.}},
\bauthor{\bsnm{Abecasis},~\bfnm{Goncalo}\binits{G.}},
\bauthor{\bsnm{Durbin}, \bfnm{Richard}\binits{R.}} \AND
\bauthor{\bsnm{{1000 Genome Project Data Processing Subgroup}}}
(\byear{2009}).
\btitle{The Sequence Alignment/Map format and SAMtools}.
\bjournal{Bioinformatics}
\bvolume{25}
\bpages{2078--2079}.
\bid{doi={10.1093/bioinformatics/btp352}, issn={1367-4811},
pii={btp352}, pmcid={2723002}, pmid={19505943}}
\end{barticle}
%
\bptok{imsref}%
\endbibitem

\bibitem{pbdR2012}
%
\begin{bmisc}[auto:STB|2014/05/27|07:42:02]
\bauthor{\bsnm{Ostrouchov},~\bfnm{G.}\binits{G.}},
\bauthor{\bsnm{Chen},~\bfnm{W.-C.}\binits{W.-C.}},
\bauthor{\bsnm{Schmidt},~\bfnm{D.}\binits{D.}} \AND
\bauthor{\bsnm{Patel},~\bfnm{P.}\binits{P.}}
(\byear{2012}).
\bhowpublished{Programming with big data in R. Available~at \url{http://r-pbd.org/}.}
\end{bmisc}
%
\bptok{imsref}%
\endbibitem

\bibitem{Biostrings}
%
\begin{bmisc}[auto:STB|2014/05/27|07:42:02]
\bauthor{\bsnm{Pag{\`e}s},~\bfnm{H.}\binits{H.}},
\bauthor{\bsnm{Aboyoun},~\bfnm{P.}\binits{P.}},
\bauthor{\bsnm{Gentleman},~\bfnm{R.}\binits{R.}} \AND
\bauthor{\bsnm{\mbox{DebRoy}},~\bfnm{S.}\binits{S.}}
(\byear{2013}).
\bhowpublished{Biostrings: String objects representing biological
sequences, and matching algorithms. R package version 2.25.6.}
\end{bmisc}
%
\bptok{imsref}%
\endbibitem


\bibitem{rman}
%
\begin{bbook}[auto:STB|2014/05/27|07:42:02]
\bauthor{R~Development~Core Team}
(\byear{2010}).
\btitle{R: A Language and Environment for Statistical Computing}.
\bpublisher{R Foundation for Statistical Computing},
\blocation{Vienna, Austria}.
\end{bbook}
%
\bptok{imsref}%
\endbibitem

\bibitem{foreach}
%
\begin{bmisc}[auto:STB|2014/05/27|07:42:02]
\bauthor{\bsnm{Revolution Analytics}} \AND
\bauthor{\bsnm{Weston},~\bfnm{Steve}\binits{S.}}
(\byear{2013}).
\bhowpublished{foreach: Foreach looping construct for R. R package
version 1.4.1}.
\end{bmisc}
%
\bptok{imsref}%
\endbibitem





\bibitem{Wickham:2010:JSSOBK:v40i01}
%
\begin{barticle}[auto:STB|2014/05/27|07:42:02]
\bauthor{\bsnm{Wickham},~\bfnm{Hadley}\binits{H.}}
(\byear{2011}).
\btitle{The split-apply-combine strategy for data analysis}.
\bjournal{Journal of Statistical Software}
\bvolume{40}
\bpages{1--29}.
\end{barticle}
%
\bptok{imsref}%
\endbibitem

\bibitem{wickham2009plumbing}
%
\begin{barticle}[mr]
\bauthor{\bsnm{Wickham},~\bfnm{Hadley}\binits{H.}},
\bauthor{\bsnm{Lawrence},~\bfnm{Michael}\binits{M.}},
\bauthor{\bsnm{Cook},~\bfnm{Dianne}\binits{D.}},
\bauthor{\bsnm{Buja},~\bfnm{Andreas}\binits{A.}},
\bauthor{\bsnm{Hofmann},~\bfnm{Heike}\binits{H.}} \AND
\bauthor{\bsnm{Swayne},~\bfnm{Deborah~F.}\binits{D.~F.}}
(\byear{2009}).
\btitle{The\vadjust{\vfill\eject} plumbing of interactive graphics}.
\bjournal{Comput. Statist.}
\bvolume{24}
\bpages{207--215}.
\bid{doi={10.1007/s00180-008-0116-x}, issn={0943-4062}, mr={2506079}}
\end{barticle}
%
\bptok{imsref}%
\endbibitem


\bibitem{biovizBase}
%
\begin{bmisc}[auto:STB|2014/05/27|07:42:02]
\bauthor{\bsnm{Yin},~\bfnm{Tengfei}\binits{T.}},
\bauthor{\bsnm{Lawrence},~\bfnm{Michael}\binits{M.}} \AND
\bauthor{\bsnm{Cook},~\bfnm{Dianne}\binits{D.}}
(\byear{2013}).
\bhowpublished{biovizBase: Basic graphic utilities for visualization of
genomic data. R~package version 1.9.1}.
\end{bmisc}
%
\bptok{imsref}%
\endbibitem

\end{thebibliography}
\end{document}